\documentclass[conference]{IEEEtran}
\IEEEoverridecommandlockouts
\usepackage[noadjust]{cite}
\usepackage[ruled,linesnumbered]{algorithm2e}
\usepackage{amsmath,amssymb,amsfonts}
\usepackage{graphicx}
\usepackage{textcomp}
\usepackage{array}
\usepackage[makeroom]{cancel}

\usepackage{mathtools}

\newcolumntype{C}[1]{>{\centering\let\newline\\\arraybackslash\hspace{0pt}}m{#1}}

\usepackage{diagbox}

\def\BibTeX{{\rm B\kern-.05em{\sc i\kern-.025em b}\kern-.08em
    T\kern-.1667em\lower.7ex\hbox{E}\kern-.125emX}}
\begin{document}

\IEEEpubid{\makebox[\columnwidth]{\textbf{978-1-5386-6163-5/18/\$31.00 \copyright 2018 IEEE} \hfill} \hspace{\columnsep}\makebox[\columnwidth]{ }}

\title{Blockchain based secure data handover scheme in non-orthogonal multiple access}


\author{\IEEEauthorblockN{Anik Islam\IEEEauthorrefmark{1},
Mohammed Belal Uddin\IEEEauthorrefmark{2}, Md. Fazlul Kader\IEEEauthorrefmark{3} and
Soo Young Shin\IEEEauthorrefmark{4}}
\IEEEauthorblockA{\IEEEauthorrefmark{1}\IEEEauthorrefmark{2}\IEEEauthorrefmark{4}Wireless and Emerging Networks System (WENS) Laboratory, Department of IT Convergence Engineering,\\
Kumoh National Institute of Technology (KIT), Gumi, South Korea\\
\IEEEauthorrefmark{3}Department of Electrical and Electronic Engineering, University of Chittagong, Chittagong, Bangladesh\\
Email: \{\IEEEauthorrefmark{1}anik.islam, \IEEEauthorrefmark{2}ahad.belal, \IEEEauthorrefmark{4}wdragon\}@kumoh.ac.kr,
\IEEEauthorrefmark{3}f.kader@cu.ac.bd}}

\maketitle

\begin{abstract}
Non-orthogonal multiple access (NOMA) with successive interference cancellation receiver is considered as one of the most potent multiple access techniques to be adopted in future wireless communication networks. Data security in the NOMA transmission scheme is on much attention drawing issue. Blockchain is a distributed peer-to-peer network enables a way of protecting information from unauthorized access, tempering etc. By utilizing encryption techniques of blockchain, a secured data communication scheme using blockchain in NOMA is proposed in this paper.  A two-phase encryption technique with key generation using different parameter is proposed. In the first-phase data is encrypted by imposing users' public key and in the second phase, a private key of the base station (BS) is engaged for encryption. Finally,  the superiority of the proposed scheme over existing scheme is proven through a comparative study based on the different features.
\end{abstract}


\section{Introduction}

\begin{figure}
	\centering
	\includegraphics[width=0.5\textwidth]{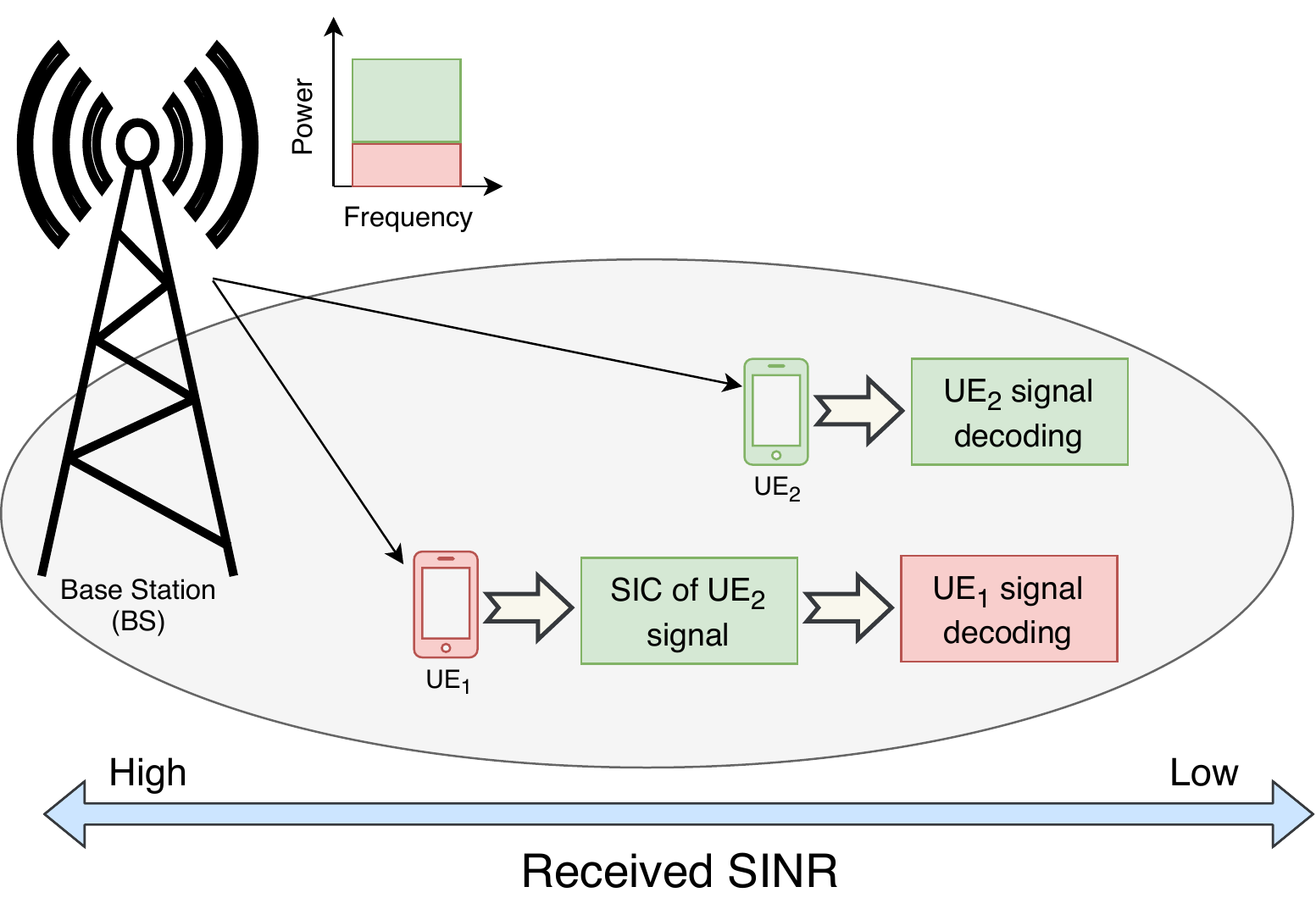}
	\caption{Basic downlink NOMA with SIC.}
	\label{fig1}
\end{figure}

\IEEEPARstart{T}{he} provision of high data rate and facilitation of multiple users to communicate simultaneously within a core network is very crucial to meet the quality of service requirement in the era of evolving wireless communication technologies. Simultaneous information exchange among a large number of devices by exploiting limited bandwidth is another challenge of future wireless communication. To overcome the problems and meet the challenges of upcoming wireless networks, non-orthogonal multiple access (NOMA) technique with successive interference cancellation (SIC) receiver is considered as one of the most promising multiple access techniques ~\cite{1,2,3,4}. In NOMA, multiple users are facilitated to transmit/receive data simultaneously at the same frequency by using power division multiplexing. In downlink NOMA, a base station (BS) transmits superposed data for the intended users by allocating more power to the weaker user UE$_2$ (User Equipment)  data than stronger user UE$_1$ data, as shown in Fig. \ref{fig1}. UE$_1$ first decodes data bits of UE$_2$. After reconstructing the signal related to those decoded data bits, it's been cancelled from the total received signal. This traditional successive interference cancellation continues till the decoding of UE$_1$'s data bits.

As described earlier, due to the way of traditional SIC procedure, the data of the weak user is not secured to the strong user. Moreover, most of the existing NOMA-based systems mainly focus on traditional SIC~\cite{1,2,3,4,5,6}. That is why proper security is needed during SIC to prevent leakage of weak users' data information to the strong user. 

\begin{figure*}[h]
	\centering
	\includegraphics[width=1\textwidth]{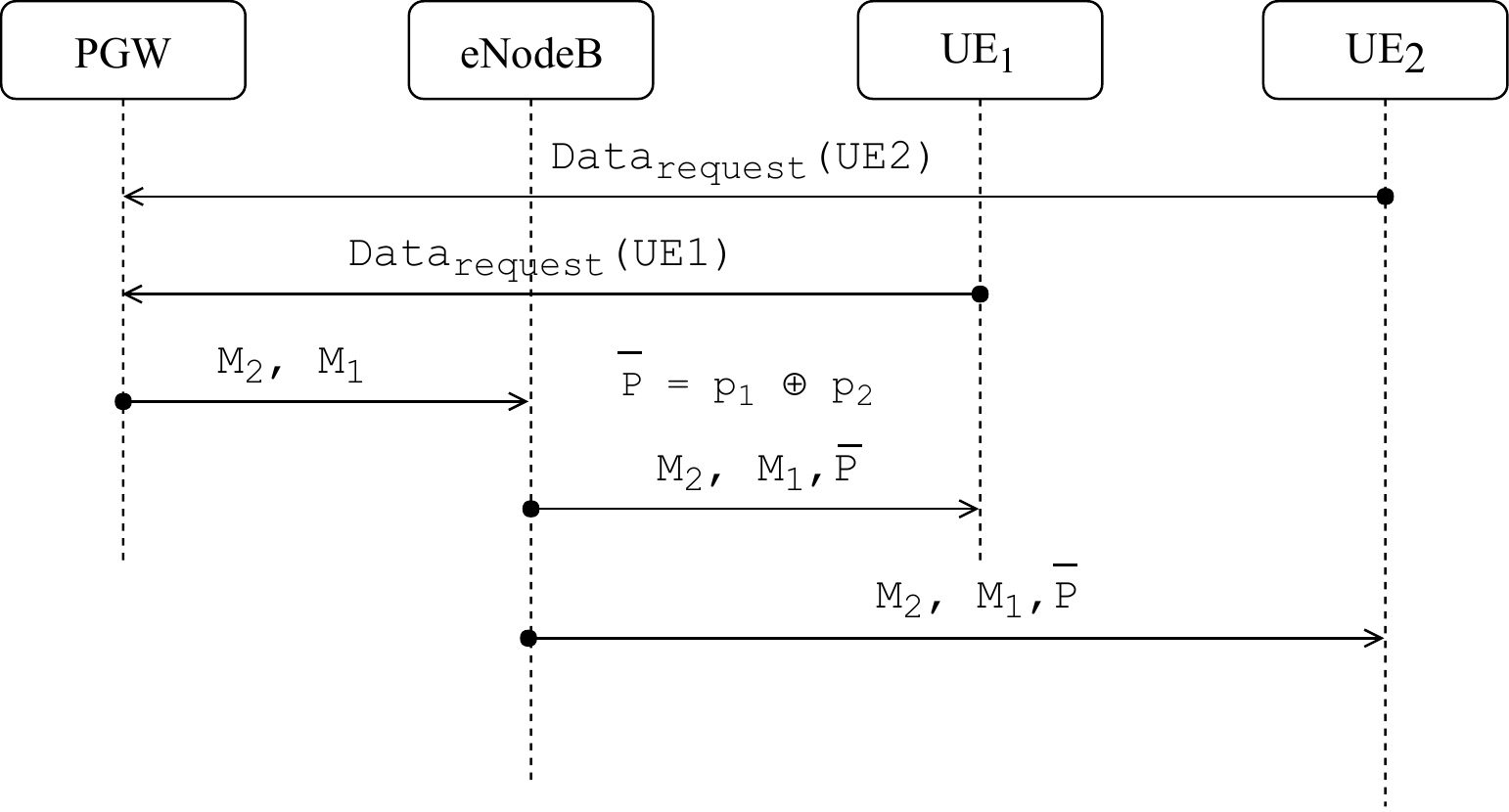}
	\caption{Existing data handover scheme~\cite{7}.}
	\label{fig2}
\end{figure*}

A secure way of performing SIC for NOMA was discussed in ~\cite{7}. In~\cite{7}, they have shown similar concern about aforementioned security issues. In order to solve that, they have adopted symmetric key encryption and they have generated the key based on international mobile equipment identity (IMEI), media access control (MAC) address. Using IMEI and MAC address could pose vulnerability towards spoofing attack. They have tried to solve it using "sticky MAC address". But this technique is used only by the vendors named CISCO and Juniper. So, this could not give solutions to the aforementioned security issues for all users. 

Blockchain has recently gained tremendous attention due to its promise of ensuring security of data. Blockchain is a data structure which is shared and replicated among the participant of the network. Blockchain was first introduced by Satoshi Nakamoto with bitcoin~\cite{8}. In blockchain, a pair of keys (private/public) is adopted. The public key is used as an identity for the user so that users' privacy can remain concealed and the private key is used for encrypting data so that data can remain protected~\cite{9,10}. However, this privacy protection technique can mitigate the issue of disclosing data to the strongest user.  

In this paper, a blockchain based secure data handover technique in NOMA is proposed in order to mitigate the mentioned above-mentioned security issues. The contribution of this paper is outlined below: 

\begin{itemize}
\item A key generation technique is proposed using different parameter.
\item A two-phase encryption is proposed so that data can be protected from any kind of attacks.
\end{itemize} 

The remaining sections of this paper are organized as follows: Section \ref{sec2} illustrates the existing data handover scheme. The proposed blockchain based data handover scheme is portrayed in Section \ref{sec3}. A performance comparison between proposed scheme and \cite{7} is demonstrated in Section \ref{sec4}. Finally, Section \ref{sec5} draws a conclusion from this paper with future research directions. 
\begin{figure*}[ht]
	\centering
	\includegraphics[width=1\textwidth]{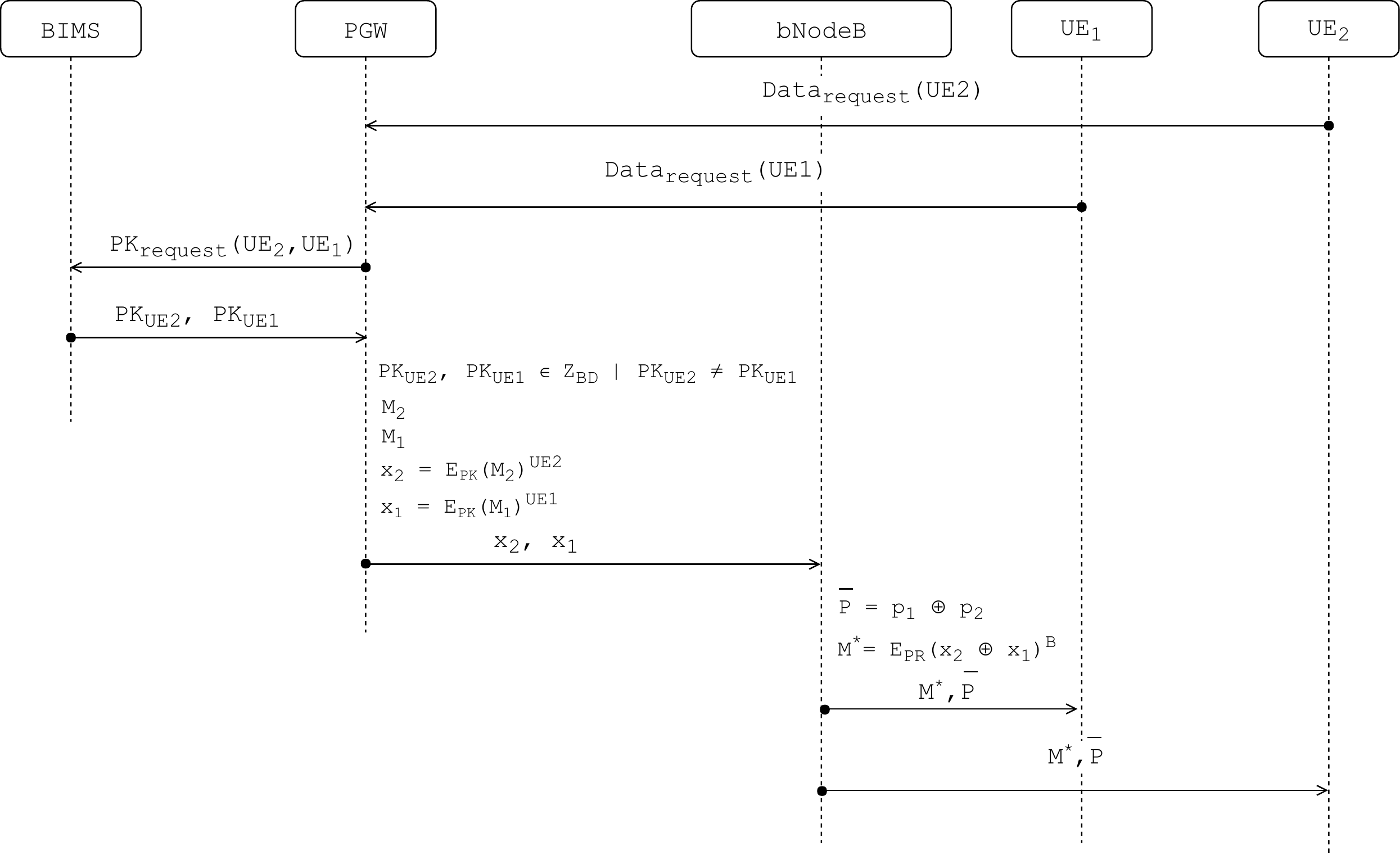}
	\caption{Proposed data handover scheme.}
	\label{fig3}
\end{figure*}

\section{Existing Data handover Scheme}
\label{sec2}

Existing data handover scheme contains UEs, eNodeB (Evolved Node-B), and PGW (Packet gateway), as shown in Fig. \ref{fig2}. All the notations and their description are provided in Table \ref{tab1}. 
{\renewcommand{\arraystretch}{1.2}
	\begin{table}[h!]
		\caption{ Notations and their description}
		\centering
		\begin{tabular}{c|l} 
			\hline
			Notation & Description \\ [0.5ex] 
			\hline\hline
			$PK_{\text{UE}_{th}}$ & Public key for UE$_{th}$  \\
			$PR_{\text{UE}_{th}}$ & Private key for UE$_{th}$  \\
			$PR_{B}$ & Private key for BS  \\
			$Z_{BD}$& Blockchain based database \\
			$p_{th}$ & Allocated power for UE$_{th}$\\ 
			$\overline{P}$ & Total allocated power \\
			$M_{th}$ & Signals for UE$_{th}$  \\			
			$E_{PK}(.)$ & Encryption using public key  \\
			$E_{PR}(.)$ & Encryption using private key  \\
			$D_{PK}(.)$ & Decryption using public key \\
			$D_{PR}(.)$ & Decryption using private key \\
			$IM$ & IMEI \\	
			$M$ & MAC address \\	
			$T$ & Timestamp \\			
			$Lat$ & Latitude \\
			$Lon$ & Longitude \\ [1ex] 
			\hline
		\end{tabular}		
		\label{tab1}
	\end{table}	
Fig. \ref{fig2} illustrates data handover scheme assuming two UEs, one BS, one PGW and the scheme is considered for downlink transmission. The procedure of existing handover applying SIC is depicted below~\cite{7}: 

\begin{enumerate}
\item Both UE$_1$ and UE$_2$ send data request to eNodeB which eNodeB forwards to PGW.
\item PGW picks unencrypted $M_2$ and $M_1$ for UE$_2$ and UE$_1$ respectively, and forwards to eNodeB. 
\item After getting $M_2$ and $M_1$ from PGW, eNodeB allocates power $p_1$ and $p_2$ with signals of UE$_2$ and UE$_1$ respectively, and superposes both signals to $\overline{P}= p_1 \oplus p_2$. eNodeB returns $M_2$ and $M_1$ with $\overline{P}$ back to UE$_2$ and UE$_1$. 
\item When UE$_1$ receives data from eNodeB, it decodes all data bits subsequently and after decoding, it subtracts signals of UE$_2$ in order to extract its own data. After that, it retrieves its own data by computing from $M_1$.

\item UE$_2$ extract its data considering signal of UE$_1$ as a noise.
\end{enumerate}

As the existing scheme does not contain any encryption techniques, data becomes vulnerable to attack. Since UE$_1$ decodes UE$_2$'s data, UE$_1$ may use UE$_2$'s data for unethical purposes. 


\section{Proposed Blockchain based data handover}
\label{sec3}

We have proposed a secure data handover process in NOMA with the integration of Blockchain. The proposed scheme contains three parts such as (1) public/private key generation, (2) encryption process in BS,  and (3) decryption process in UE.

\subsection{Public/private key generation}

In the proposed scheme, private/public keys play a very significant role. These keys are used for first-phase encryption. The proposed scheme assumes a protected area in UE, named Kaladanda Box (KBox), which preserves credentials like UE's private/public key, BS's private key. An assumption is made that when UE registers in the BS, BS shares its private key $PR_B$ with every registered UE. $PR_B$ uses for second-phase encryption. However, before registering in BS, UE generates the public key and broadcast it for secure communication. In order to construct the public key, a private key is needed. Let $PR_{\text{UE}_i}$ is a private key of $i^{th}$ UE. 

\[  PR_{\text{UE}_i} = SHA256(IM_i, M_i, T_i, Lat_i, Lon_i)\] 

Here, $i \in \{1, 2,...., n \}$ and $n$ is total UE and for our proposed scheme $n=2$ and let $\overline{PR}_{\text{UE}}$ is a set of UE's private key.

\[ \overline{PR}_{\text{UE}} = \{  PR_{\text{UE}_1}, PR_{\text{UE}_2},.....,PR_{\text{UE}_n} \}_{\neq} \]

Let $G$ is a set of $(x, y)$ coordinates on the elliptic curve. 

\[  PK_{\text{UE}_i} =  PR_{\text{UE}_i} \otimes (G_x, G_y)  \] 

After generating the public key, UE broadcasts, and saves public key KBox for decrypting data bits.

\subsection{Encryption process in BS}

In this part, a two-phase encryption is adopted in order to give protection against different kinds of attacks, as shown in Fig. ~\ref{fig3}. In the first phase, an asymmetric encryption technique is adopted and data is encrypted by UE's public key so that only authorized UE can decrypt it using the private key. In the second phase, a symmetric encryption technique is adopted and data again encrypted using BS's private key so that only legitimate can access data and become safe from eavesdroppers. We have assumed a modified version of eNodeB named blockchain supported Node-B (bNodeB) which supports blockchain integration in the BS. First, both UE$_2$ and UE$_1$ send data requests to bNodeB by sending Data$_{\text{request}}$(UE$_2$) and Data$_{\text{request}}$(UE$_1$) respectively. bNodeB forwards the request to PGW to process the request. When PGW gets requests from UE$_2$ and UE$_1$, PGW retrieves public keys of both UE$_2$ and UE$_1$ from  blockchain based identity management server (BIMS) by sending PK$_{\text{request}}$(UE$_2$, UE$_1$). BIMS responds with PK$_{\text{UE2}}$ and PK$_{\text{UE1}}$. Let $Z_{BD}$ is the blockchain based database that manages public keys. 

\[\text{PK}_\text{UE2},\text{ } \text{PK}_\text{UE1} \in Z_{BD} \text{ }| \text{ } \text{PK}_\text{UE2} \neq  \text{PK}_\text{UE1} \]

As PK is used both as an identity and as securing data packet, each user's PK should be unique. In order to make PK unique,  we choose  not only unique IMEI, MAC address, but also timestamp and spatial information for generating users' key. The key becomes strong after mixing timestamp and spatial information because if any hacker tries to clone IMEI or spoof the MAC address, he still needs timestamp and spatial information to clone private/public key. However, PGW picks $M_2$ and $M_1$ for UE$_2$ and UE$_1$ respectively. Let $x_2$ and $x_1$ are encrypted packets.

\[ x_2 = E_{PK}(M_2)^{UE2} \]
\[ x_1 = E_{PK}(M_1)^{UE1} \]

PGW forwards these encrypted data $x_2$ and $x_1$ to bNodeB for further processing. As bNodeB receives $x_2$ and $x_1$, bNodeB allocates power $p_2$ and $p_1$ with signals of UE$_2$ and UE$_1$ respectively, and superposes both signals to $\overline{P}= p_1 \oplus p_2$. After that, both $x_2$ and $x_1$ experience encryption by employing $PR_B$. Let $M^*$ is encrypted packets.

\[ M^*= E_{PR}(x_2 \oplus x_1)^B \]

Here, only legitimate users hold $PR_B$.

\subsection{Decryption process in UE}

In this part, a two-phase decryption process is discussed in Algorithm \ref{alg1}. Upon receiving a response from  bNodeB, both UE$_1$ and UE$_2$ first try to decrypt received packets $D_{PR}(M^*)^B$ using $PR_B$. As both UE$_1$ and UE$_2$ is legitimate users, both UE$_1$ and UE$_2$ contains $PR_B$ in its KBox. After that, UE$_1$ subtracts  UE$_2$'s data after decoding all the bits sequentially in order to extract its own data. As data for both UE$_2$ and UE$_1$ is encrypted by their public key, UE$_1$ requires UE$_2$'s private key in order to read UE$_2$'s data. So, UE$_2$'s data remains protected during decoding. However, UE$_1$ decrypts its data $D_{PR}(M_1)^{\text{UE1}}$ using $PR_{\text{UE1}}$.  After the decryption process, UE$_1$ retrieves its own data by computing from $\overline{dbm}$. Following this, UE$_2$ decrypts data $D_{PR}(M^*)^B$ using $PR_B$. UE$_2$ extracts its own data considering UE$_1$'s data as a noise. After extracting data, UE$_2$ decrypts data $D_{PR}(M_2)^{\text{UE2}}$ using $PR_{\text{UE2}}$. Finally, UE$_2$ retrieves its own data by computing from $\overline{dbm}$.

 \begin{algorithm}[htb]
\SetAlgoLined
\SetAlgoNoLine
\DontPrintSemicolon
$\overline {UE}$ : set of resgitered user.\; 
$M^*$ : encrypted data bits from BS.\;
$\overline{b} = D_{PR}(M^*)^B $.\;
\eIf{ \text{UE}$_1 \in \overline {UE} $ }{ 
      \While{$ b \in \overline{b}$}{
             $db = decode(b)$.\;
             \eIf{$db \notin  \text{UE}_1$}{
                  $\cancel{db}$.\;
              }{
                  $db \xRightarrow[]{Include} M_1$.\;
             }
        }
     $\overline{dbm} = D_{PR}(M_1)^{\text{UE1}} $.\;
   }{
      \While{$ b \in \overline{b}$}{
             $db = decode(b)$.\;
             \eIf{$db \notin  \text{UE}_2$}{
                  $\cancel{db}$.\;
              }{
                $db \xRightarrow[]{Include} M_2$.\;
               }
        }
     $\overline{dbm} = D_{PR}(M_2)^{\text{UE2}} $.\;
   }
 \caption{Two-phase decryption process in UE}
 \label{alg1}
\end{algorithm}

%
%
%
%
%


\section{Performance Analysis}

\label{sec4}

A performance comparison is outlined in Table \ref{table2}.  The features that take into consideration for performance analysis are delineated below: 

\begin{description}
  \item[User privacy] \hfill \\ This feature protects the user personal information from leaking while registering in BS. The proposed scheme has taken into consideration on the issue and proposed to share information in minimum level which is managed in blockchain.   
  \item[Encryption] \hfill \\ This feature protects data from unauthorized access. Both proposed scheme and secure SIC~\cite{7} have introduced two-phase encryption. The proposed scheme has adopted asymmetric encryption using the public key in the first phase and symmetric encryption using the private key of BS in the second phase. On the contrary,  secure SIC~\cite{7} has used symmetric encryption in both phases. 
  \item[Key generation] \hfill \\ This feature covers comparison of properties that are used for key generation. Secure SIC~\cite{7} has utilized IMEI and MAC address for generating the key. On the contrary, the proposed scheme has taken not only IMEI and MAC address into consideration, but also has taken timestamp and spatial information into consideration. The key generation for the attacker is much more difficult in the proposed scheme than secure SIC~\cite{7}.
  \item[Protection against spoofing attack] \hfill \\ This feature supports the protection against spoofing attack. Both proposed and secure SIC~\cite{7} has proposed protection against spoofing attack. However, the solution in secure SIC~\cite{7} only covers two vendors named CISCO and Juniper. But the solution in the proposed schemes supports every user.  
  \item[Protection against data hijacking] \hfill \\ This feature supports data protection against unauthorized access. Both the proposed scheme and secure SIC~\cite{7} have provided their solution against hijacking data by employing encryption.
\end{description}

\begin{table}[h]
\caption{Performance comparison between proposed scheme and Secure SIC~\cite{7}}
\label{table2}	
\centering
 \begin{tabular}{| C{2.85cm} | C{2.1cm} | C{2.1cm} |  } 
 \hline
\diagbox[innerwidth=2.85cm,innerleftsep=1cm]{ Features} {Proposed} & Proposed scheme &  Secure SIC~\cite{7} \\ [0.5ex]
 \hline
User privacy  & yes & no  \\
 \hline
Encryption   & yes & yes  \\
 \hline
Key generation   & IMEI, MAC address, timestamp, spatial information & IMEI, MAC address  \\
 \hline
Protection against spoofing attack  & for all & partial (only for CISCO and Juniper) \\
 \hline
Protection against data hijacking  & yes & yes \\ 
\hline

 \end{tabular}
\end{table}

\section{Conclusion}
\label{sec5}
In this paper, we have proposed a secure data handover scheme combining with blockchain. In the proposed scheme, UE generates a private key based on IMEI, MAC address, timestamp, Lat, and Lon. After that UE generates public key out of private key and shared with BS and also BS shares its private key to the registered users so that only legitimate users can access transmitted information. Complexity analysis of adopting both symmetric and asymmetric encryption in NOMA along with secrecy analysis is kept for future extension of this paper. However, the detailed discussion regarding UE's identity management using blockchain needs to be researched which can be subjected to future works.

\section*{Acknowledgment}

This work was supported by the Brain Korea 21 Plus Project (Department of IT Convergence Engineering, Kumoh National Institute of Technology).

\bibliographystyle{IEEEtran}
\bibliography{IEEEabrv,SecureSIC}

\end{document}